\documentclass[12pt]{iopart}
\usepackage{graphics}
\usepackage{epsfig}
\usepackage{citesort}

\begin{document}

\newcommand{\gtrsim}{\,\rlap{\lower 3.5 pt \hbox{$\mathchar \sim$}} \raise
1pt \hbox {$>$}\,}
\newcommand{\lesssim}{\,\rlap{\lower 3.5 pt \hbox{$\mathchar \sim$}} \raise
1pt \hbox {$<$}\,}

\title{Reconstructing the primordial power spectrum - a new algorithm}
\author{Steen Hannestad}
\address{Department of Physics, University of Southern Denmark,
Campusvej 55, DK-5230 Odense M, Denmark \\ and \\
NORDITA, Blegdamsvej 17, DK-2100 Copenhagen, Denmark}

\date{{\today}}

\begin{abstract}
We propose an efficient and model independent method for
reconstructing the primordial power spectrum from Cosmic Microwave
Background (CMB) and large scale structure observations. The
algorithm is based on a Monte Carlo principle and therefore very
simple to incorporate into existing codes such as Markov Chain
Monte Carlo.

The algorithm has been used on present cosmological data to test
for features in the primordial power spectrum. No significant
evidence for features is found, although there is a slight
preference for an overall bending of the spectrum, as well as a
decrease in power at very large scales.

We have also tested the algorithm on mock high precision CMB data,
calculated from models with non-scale invariant primordial
spectra. The algorithm efficiently extracts the underlying
spectrum, as well as the other cosmological parameters in each
case.

Finally we have used the algorithm on a model where an artificial
glitch in the CMB spectrum has been imposed, like the ones seen in
the WMAP data. In this case it is found that, although the
underlying cosmological parameters can be extracted, the recovered
power spectrum can show significant spurious features, such as
bending, even if the true spectrum is scale invariant.
\end{abstract}
\maketitle

\section{Introduction}

The present cosmological concordance model has been shown to
describe all present data surprisingly well with only about six
parameters \cite{Tegmark:2003ud,map2}. From a physics point of
view it is based on reasonable assumptions, such as general
relativity and forms of matter with constant equations of state,
albeit the physical nature of both the dark matter and the dark
energy remains unknown. The initial perturbation spectrum appears
to be almost purely adiabatic, Gaussian and scale invariant.
Furthermore a general framework for producing such fluctuations in
the early universe, namely inflation, is known and well
established. As long as the universe remains deeply within the
slow-roll regime during inflation, and the inflaton remains weakly
coupled, Gaussian quantum fluctuations are produced which
eventually become the adiabatic metric fluctuations observed for
instance in the cosmic microwave background. However, there are
many reasons to believe that there are deviations from this
behaviour during the inflationary epoch, and countless physical
mechanisms have been investigated. In the simplest models the
deviations from scale invariance are caused by non-zero
derivatives of the inflaton potential. Depending on the exact
inflation model this can produce a tilt in the spectrum, and in
general one can expand the produced spectrum of metric
fluctuations in a Taylor series around the scale-invariant
Harrison-Zeldovich spectrum

\begin{equation}
\log P (k) = \log P (k_0) + (n-1) \log \frac{k}{k_0} +\left.
\frac{1}{2} \frac{d\, n}{d \, \log k} \right|_{k_0} \, \log ^2
\frac{k}{k_0} + \cdots \label{power}
\end{equation}

where the different terms in the expansion are related to the
derivatives of the potential.

Within this framework it has to be assumed that the universe does not
deviate significantly from slow-roll in the sense that the power series
above converges, and higher order terms are very small.

However, the present cosmological data shows some indication that
the second order term is not small compared with the first order
term
\cite{map3,Leach:2003us,Lue:2003su,Barger:2003ym,Kinney:2003uw,Kinney:2000nc,Hannestad:2000tj,Hannestad:2001nu},
perhaps indicating that the above series expansion does not
converge, and that slow-roll may not be a valid approximation.

If one does away with the assumption of slow-roll during the
production of fluctuations then much more radical deviations from
scale invariance can be achieved. If for instance there are phase
transitions or resonant production of particles during inflation,
then step-like features in the power spectrum can occur
\cite{Starobinsky:ts,Chung:1999ve,Elgaroy:2003hp,Barriga:2000nk,Adams:1997de}.
It is in fact possible that such a feature is seen in the CMB data
at the very largest scales, where the level of fluctuations is
significantly smaller than expected.

On a different note it is also possible that effects from Planck
scale physics can show up during inflationary slow-roll which
occurs at much lower energies and perturb the spectrum of
fluctuations
\cite{Martin:2000xs,Brandenberger:2000wr,Niemeyer:2000eh,Easther:2001fi,Easther:2001fz,Danielsson:2002kx,Easther:2002xe,Alberghi:2003am,Martin:2003kp,Bergstrom:2002yd,Elgaroy:2003gq,Martin:2003sg},
one possibility being a logarithmic modulation of the power
spectrum. This can occur even if the slow-roll conditions are
fulfilled.

It is of course possible to search for specific types of features
predicted in a given type of model, such as steps, Gaussian peaks,
or oscillations (see e.g.\ \cite{Elgaroy:2001wu}). However, with
such a plethora of different possibilities it would be highly
desirable with a model-independent and robust method for searching
for deviations from scale invariance in the power spectrum. Such a
method should be able to naturally encompass the power-law spectra
predicted by standard slow-roll, as well as the more radical
features such as steps and bumps.

In the present paper we describe exactly such a method which is able
to extract features in the underlying spectrum, using either CMB
or large scale structure data. Furthermore it is very easy to implement
in existing software for cosmological parameter estimation.

In section 2 we describe general features of likelihood analysis,
as well as the present cosmological data. In sections 3 and 4 we
describe the spectrum reconstruction algorithm and use it on
present cosmological data to search for features. In section 5 we
go on to test the algorithm against mock data from future CMB
experiments such as the Planck Surveyor. In section 6 we test how
systematic errors in the $C_l$ data can affect parameter
estimation and spectrum recovery. To that end we insert a Gaussian
feature directly into the mock CMB data and do likelihood analysis
on this modified data. Finally section 7 contains a discussion and
conclusion.


\section{Likelihood analysis}

For calculating the theoretical CMB and matter power spectra we
use the publicly available CMBFAST package \cite{CMBFAST}. As the
set of cosmological parameters other than those related to the
primordial spectrum we choose the minimum standard model with 4
parameters: $\Omega_m$, the matter density, the curvature
parameter, $\Omega_b$, the baryon density, $H_0$, the Hubble
parameter, and $\tau$, the optical depth to reionization. We
restrict the analysis to geometrically flat models $\Omega =
\Omega_m + \Omega_\Lambda = 1$. When using large scale structure
data we use the bias parameter, $b$, as a free parameter, which is
a conservative, but well motivated choice.

\begin{table}
\begin{center}
\begin{tabular}{|l|c|}
\hline
parameter & prior\cr
\hline
$\Omega_m$ & $0.28 \pm 0.14$ (Gaussian) \cr
$h$ & $0.72 \pm 0.08$ (Gaussian) \cr
$\Omega_b h^2$ & $0.014-0.040$ (Top Hat) \cr
$\tau$ & $0-1$ (Top hat) \cr
$b$ & free \cr
\hline
\end{tabular}
\end{center}
\caption{The different priors on parameters
used in the likelihood analysis. Parameters related to the power spectrum
are not tabulated here.}
\label{table:prior}
\end{table}

In this numerical likelihood analysis we use the free parameters
discussed above with certain priors determined from cosmological
observations other than CMB and LSS. In flat models the matter
density is restricted by observations of Type Ia supernovae to be
$\Omega_m = 0.28 \pm 0.14$ \cite{Perlmutter:1998np}. Furthermore,
the HST Hubble key project has obtained a constraint on $H_0$ of
$72 \pm 8 \,\, {\rm km} \, {\rm s}^{-1} \, {\rm Mpc}^{-1}$
\cite{freedman}. A summary of the priors can be found in table 1.
The actual marginalization over parameters was performed using a
simulated annealing procedure \cite{Hannestad:wx}.

\subsection{Cosmological data}

{\it Large scale structure --} At present there are two large
galaxy surveys of comparable size, the Sloan Digital Sky Survey
(SDSS) \cite{Tegmark:2003uf,Tegmark:2003ud}
 and the 2dFGRS (2 degree Field Galaxy Redshift Survey)\cite{2dFGRS}
Once the SDSS is completed in 2005 it will be significantly larger
and more accurate than the 2dFGRS. At present the two surveys are,
however, comparable in precision and in the present paper we use
data from the 2dFGRS.

Tegmark, Hamilton and Xu \cite{THX} have calculated a power
spectrum, $P(k)$, from this data, which we use in the present
work. The 2dFGRS data extends to very small scales where there are
large effects of non-linearity. Since we only calculate linear
power spectra, we use (in accordance with standard procedure) only
data on scales larger than $k = 0.2 h \,\, {\rm Mpc}^{-1}$, where
effects of non-linearity should be minimal (see for instance
Ref.~\cite{Tegmark:2003ud} for a discussion). Making this cut
reduces the number of power spectrum data points to 18.

{\it CMB --}
The CMB temperature
fluctuations are conveniently described in terms of the
spherical harmonics power spectrum
$C_l \equiv \langle |a_{lm}|^2 \rangle$,
where
$\frac{\Delta T}{T} (\theta,\phi) = \sum_{lm} a_{lm}Y_{lm}(\theta,\phi)$.
Since Thomson scattering polarizes light there are also power spectra
coming from the polarization. The polarization can be
divided into a curl-free $(E)$ and a curl $(B)$ component, yielding
four independent power spectra: $C_{T,l}, C_{E,l}, C_{B,l}$ and
the temperature $E$-polarization cross-correlation $C_{TE,l}$.

The WMAP experiment have reported data only on $C_{T,l}$ and $C_{TE,l}$,
as described in Ref.~\cite{map1,map2,map3,map4}

We have performed the likelihood analysis using the prescription
given by the WMAP collaboration which includes the correlation
between different $C_l$'s \cite{map1,map2,map3,map4}. Foreground contamination has
already been subtracted from their published data.

In parts of the data analysis we also add other CMB data from
the compilation by Wang {\it et al.} \cite{wang3}
which includes data at high $l$.
Altogether this data set has 28 data points.


\section{Reconstructing the power spectrum in bins}

One method which has often been used to search for power spectrum
features is to bin the spectrum into $N$ bins in $k$-space and
then calculate what could be called a band amplitude in each of
the bins \cite{Wang:1998gb,Hannestad:2000pm}. This is the crudest
possible model and does not yield a continuous power spectrum.
While there are models prediction such features they do not
generally appear.

Another method is to do linear interpolation between the bin
amplitudes. This was for instance done in the work by Bridle et
al. \cite{Bridle:2003sa} on the WMAP+2dF data (see also
\cite{Wang:2000js}). In this paper a parametrization was used
where power at a given $k$ was
\begin{equation}
P(k) = \cases{\frac{(k_{i+1}-k)b_i+(k-k_i)b_{i+1}}{k_{i+1}-k_i} &
for $k_i < k < k_{i+1}$ \cr
b_n & for $k > k_n$},
\end{equation}
where $b_i$ are the bin amplitudes and $k_i$ the bin positions.

However, this parametrization has the disadvantage that is does not
fit a simple power-law spectrum. While the amplitude at each bin
point can fit an overall power-law, the linear interpolation between
bin points can only fit a simple power law if $n=1$.

The method we use for binning is therefore slightly different. We
choose a $k_{\rm min}$ which is smaller than the minimum used in
actual calculations (typically $k_{\rm min} = 10^{-5} \, h$/Mpc)
and $k_{\rm max}$ so that it is higher than the largest value used
(typically $k_{\rm max} = 1 \, h$/Mpc). From the number of bins
$N$ we then calculate $N+1$ points in $k$-space which are
logarithmically spaced, i.e.
\begin{equation}
k_i = k_{\rm min} \left(\frac{k_{\rm max}}{k_{\rm min}}\right)^{i/N}
\,\,\,\, , \,\,\,\, i = 0,...,N
\end{equation}
At each of these points there is a power spectrum amplitude, $A_i
= \log(P(k_i))$. From this set of $N+1$ points linear
interpolation is done in logarithmic space
\begin{equation}
A(k) = \cases{A(k_0) & $k < k_{\rm min}$ \cr
A(k_i) + (\log(k)-\log(k_i))\frac{A_{i+1}-A_i}{\log(k_{i+1})-\log(k_i)}
& $k_{\rm min} < k < k_{\rm max}$ \cr
A(k_N) & $k > k_{\rm max}$ }
\end{equation}

This particular type of interpolation has the advantage that it has
the simple power law spectrum as a special case.
With this choice of binning there are $N+1$ free parameters (amplitudes)
which can either be chosen using a Monte Carlo type algorithm or calculated
on a grid. However, for large $N$ grid based methods become unfeasible.

The question is then how to judge whether adding more bins
effectively improves the fit. $\chi^2$ in itself is not a good
measure of whether the agreement with data is improved by
increasing the number of bins because the effective number of
degrees of freedom of the system changes when the number of bins
changes. Instead we use the Goodness-of-Fit (GoF), defined as the
probability of $\chi^2$ being larger than the obtained value by
pure chance alone
\begin{equation}
{\rm GoF} = \frac{1}{\Gamma(\nu/2)}\int_{\chi^2/2}^\infty
e^{-t}t^{\nu/2-1} dt, \label{eq:gof}
\end{equation}
where $\nu$ is the effective number of degrees of freedom.

As long as the GoF increases when $N$ increases then there is a
preference in the data for a feature or slope of some kind.
Therefore the algorithm should be run with increasing $N$ until
the GoF no longer increases. When the spectrum reconstruction has
converged the GoF will actually decrease with increasing $N$
because $\nu$ increases.

Note that we do not calculate error bars on the bin amplitudes for
each $N$. Rather the analysis can effectively be seen as a
likelihood analysis in terms of a single parameter $N$, where all
other parameters are marginalized over. However, as described
above it is slightly different from a normal likelihood analysis
because the number of degrees of freedom, $\nu$, changes with $N$.

It should be noted here that there is a radically different method
for reconstructing the primordial power spectrum from CMB
observations which relies on the fact that for {\it constant}
cosmological parameters, the $C_l$ spectrum can be inverted to
find $P(k)$ \cite{Kogo:2003yb,Matsumiya:2002tx}. However, the
weakness of the method is that since any reasonable likelihood
calculation must vary the cosmological parameters to find the best
fit, spurious features can occur in the extracted power spectrum.
Another possibility is to replace the binning with some other
weighing, such as using a wavelet method
\cite{Mukherjee:2003cz,Mukherjee:2003yx,Mukherjee:2003ag}. In this
latter case one estimates the amplitudes of a given number of
basis functions. If these are chosen optimally then the method can
be very powerful in picking up localized features in the power
spectrum \cite{Mukherjee:2003yx}.

\subsection{Application to present data}

As a particular example we have used the binning method described
here on current cosmological data.

\begin{figure}
\begin{center}
\includegraphics[width=110mm]{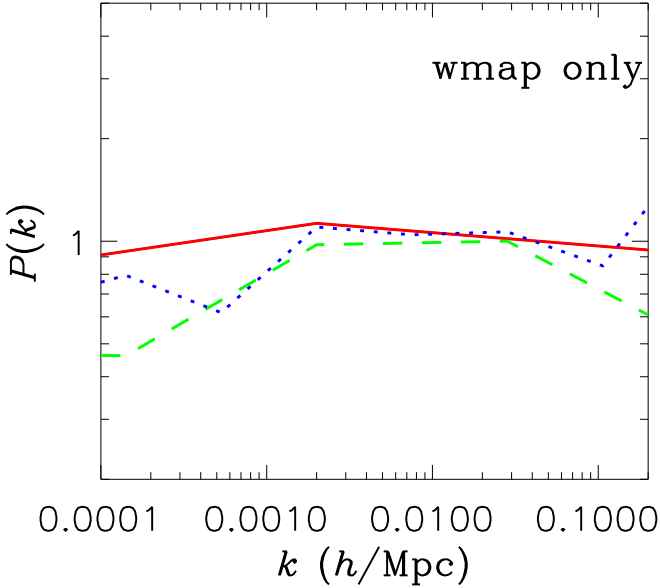}
\caption{Reconstructed best fit power spectra to the WMAP data
from the fixed bin algorithm. The full line is for $N=2$, the
dashed for $N=4$, and the dotted for $N=8$.}
\label{fig:spec1}
\end{center}
\end{figure}

\begin{figure}
\begin{center}
\includegraphics[width=110mm]{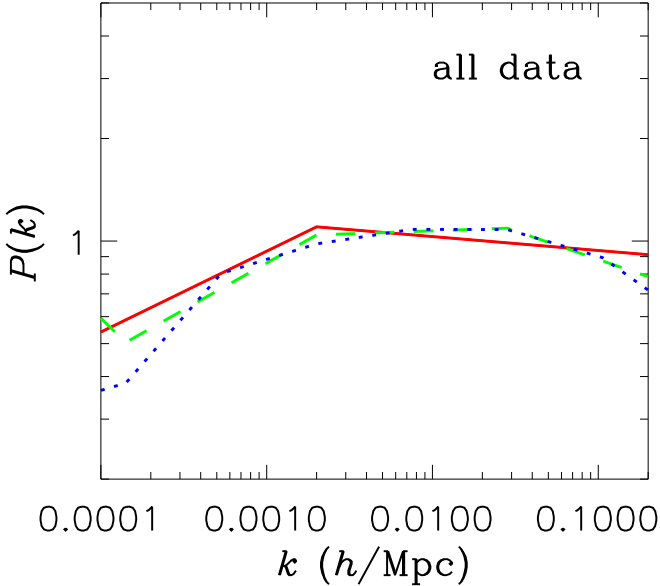}
\caption{Reconstructed best fit power spectra to all present data
from the fixed bin algorithm. The full line is for $N=2$, the
dashed for $N=4$, and the dotted for $N=8$.} \label{fig:spec2}
\end{center}
\end{figure}

We divide the likelihood analysis into two parts:
\vspace*{0.2cm}\\
a) WMAP data only (referred to as WMAP only) \\
b) WMAP + Wang et al. + 2dFGRS data (referred to as "all data")
\vspace*{0.2cm} \\
The results can be seen in Table \ref{table:fixed}, as well as in
figures \ref{fig:spec1} and \ref{fig:spec2}. From Table
\ref{table:fixed} we note that the GoF is quite poor for the WMAP
data, a fact which is well known and ascribable to the fact that
there are localized glitches in the power spectrum. While the
physical meaning of these glitches is not yet fully understood it
is likely that they have to do with the power spectrum
reconstruction from the CMB maps, and not with any fundamental
physics effect.

What is perhaps more interesting is that the application of the spectrum
reconstruction algorithm does not lead to any significant improvement
in the GoF. In most cases the $\chi^2$ value is almost unchanged.

From figures \ref{fig:spec1} and \ref{fig:spec2} it is clear,
however, that the data shows a slight preference for a negative
running of the effective spectral index, i.e.\ the spectrum has an
overall negative curvature. Exactly the same is also found in
studies of the power spectrum within the slow roll approximation.

The spectrum reconstruction also finds that there is a dip in
power around the scale of the lowest measurable multipoles. From
the rough relation $l \sim 2 k c H_0^{-1}$ (which applies exactly
only to the flat, pure CDM model), we find that there some
indication in the data for a dip around $k = 0.001 \, h$/Mpc,
corresponding to $l \sim 5-10$. This again corresponds quite well
to what is found in other studies
\cite{Cline:2003ve,Contaldi:2003zv,Feng:2003zu,Efstathiou:2003tv,Efstathiou:2003wr,deOliveira-Costa:2003pu}.

\begin{figure}
\begin{center}
\includegraphics[width=110mm]{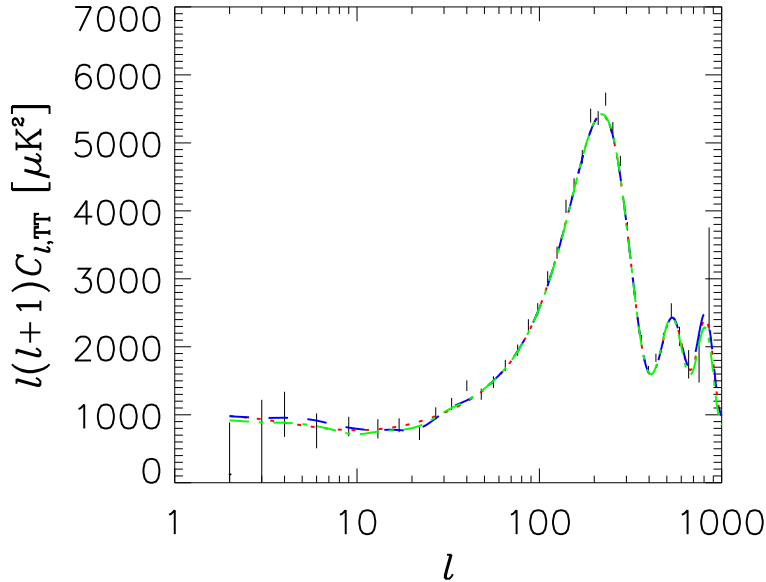}
\caption{The temperature (TT) power spectra for the best fit
models to the WMAP only data. The solid line is for $N=2$, the
dashed for $N=4$, and the dotted for $N=8$. The black lines are
the binned WMAP data.} \label{fig:cl}
\end{center}
\end{figure}

On the other hand, even though this seems like a pronounced effect
in $P(k)$, the corresponding $C_l$ spectra shown in figure
\ref{fig:cl} are almost identical at low multipoles.

\begin{figure}
\begin{center}
\includegraphics[width=75mm]{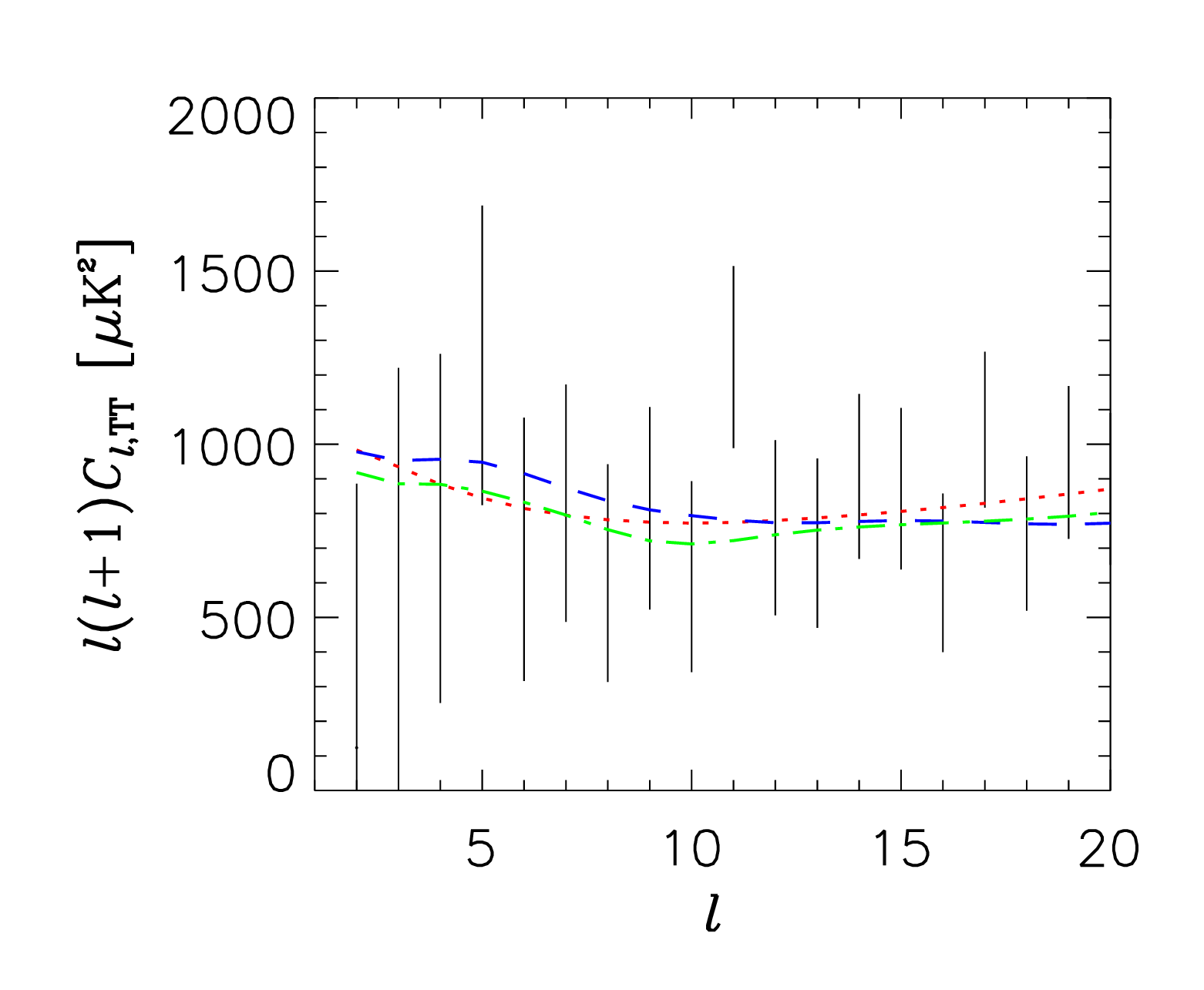}
\includegraphics[width=75mm]{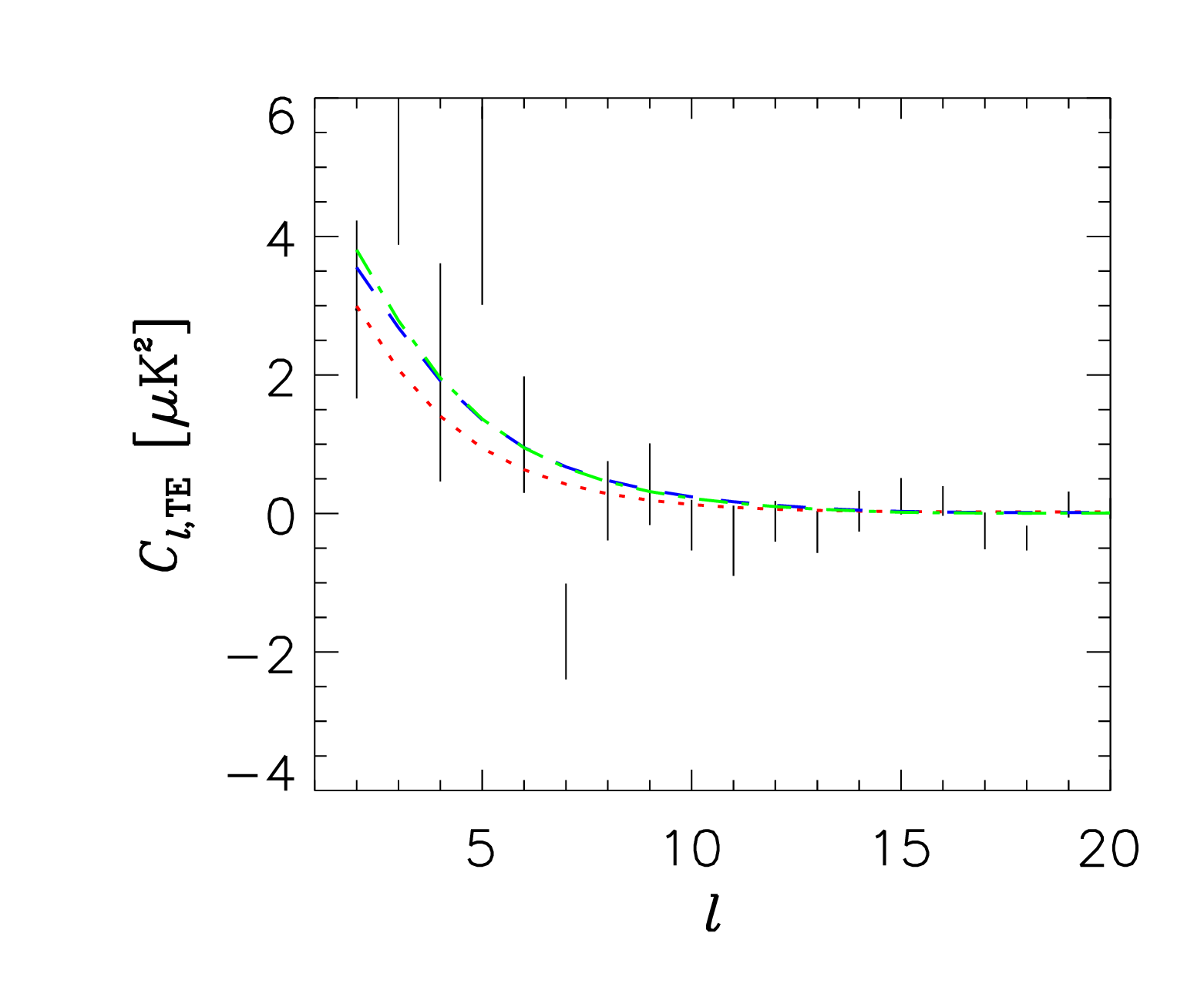}
\caption{The temperature (TT) and temperature-polarization (TE)
power spectra for the best fit models to the WMAP only data. The
solid line is for $N=2$, the dashed for $N=4$, and the dotted for
$N=8$. Only the low-$l$ parts of the spectra are shown. The black
lines are the WMAP data.} \label{fig:lowl}
\end{center}
\end{figure}

One might wonder why this is the case when the fit would obviously
be better if power was significantly reduced on small scales. The
answer is that it is not possible to lower the amplitude of the
$TT$ spectrum at small multipoles by changing the underlying power
spectrum without doing the same to the $TE$ spectrum. This on the
other hand fits well with a high amplitude at low multipoles,
something which is usually ascribed to a high optical depth. This
effect on the two different spectra at low multipoles can be seen
in figure \ref{fig:lowl}

The end result is that the overall best fit with regards to the
power spectrum is one where there is a small decrease in amplitude
at low $l$, but not sufficiently large that the $TE$ spectrum
becomes inconsistent with data.

\begin{table}
\begin{center}
\begin{tabular}{|l|ll|ll|}
\hline & \multicolumn{2}{l|}{\em WMAP data only} &
\multicolumn{2}{l|}{\em all data} \cr \hline $N$ & $\chi^2$ & GoF
& $\chi^2$ & GoF \cr \hline 2 & 1431.08 & 0.0432 & 1465.89 &
0.0742 \cr 4 & 1429.58 & 0.0422 & 1460.46 & 0.0836 \cr 8 & 1429.19
& 0.0364 & 1459.62 & 0.0745 \cr \hline
\end{tabular}
\end{center}
\caption{The smallest $\chi^2$ recovered for the fixed bin
algorithm, based on the given data set. The Goodness-of-Fit (GoF)
is defined in Eq.~(\ref{eq:gof}).} \label{table:fixed}
\end{table}


\section{The sliding bin method}

In the previous section we discussed a model independent way of
searching for power spectrum variations and used it on present cosmological
data.
However, the binning in $k$-space
was chosen a priori, a fact which can render the method almost useless
in some instances. One example is the possibility of a step-like
feature in the power spectrum. Such an abrupt feature cannot be fitted
by the algorithm presented in the previous section, unless the
number of bins is taken to be very large.

Of course one could search for steps in the spectrum by doing a
new, independent scan of parameter space using the step height and
location as free parameters. This method was for instance used by
Bridle et al. \cite{Bridle:2003sa} to search for various kinds of
spectral effects in the present cosmological data.

However, it is clearly desirable to have a robust and more general
method for searching for spectral deviations. A natural step is to
allow the binning in $k$ to be free variables. Here we propose one
such method which is simple to implement and fits extremely well
with Monte Carlo based algorithms such as MCMC
\cite{Christensen:2001gj,Lewis:2002ah} or simulated annealing
\cite{Hannestad:wx}.

{\noindent}The method goes as follows:
\begin{tabbing}
{\bf a)} \= Choose the number of $k$-space bins. This {\it must} now be a power
of 2 \\
\> (i.e. 2, 4, 8, 16, ...).\vspace*{0.2cm} \\
{\bf b)} As in the previous algorithm calculate $N+1$ amplitudes.\vspace*{0.2cm} \\
{\bf c)} Choose the lowest and highest $k$ values as fixed numbers. These should be chosen \\
\> so that $k_{\rm min}$ is smaller than the minimum used in actual calculations and $k_{\rm max}$ \\
\> so that it is higher than the largest value used. \vspace*{0.2cm} \\
{\bf d)} Start dividing up $k$-space. First choose one $k$-value, $k_1$, at random between \\
\> the starting points with a flat prior so that $k_1 = \alpha_1 (k_{\rm max} - k_{\rm min}) + k_{\rm min}$, \\
\> where $\alpha_1 \in [0,1]$ is a random number provided by for instance the MCMC \\
\> algorithm. Next, choose a $k_2$ in the same way at random between $k_{\rm min}$ and $k_1$, \\
\> and a $k_3$ between $k_1$ and $k_{\max}$. \vspace*{0.2cm} \\
{\bf e)} Continue this iterative procedure until $N-1$ values have been chosen. \\
\> This requires $\log(N)/\log(2)$ steps of the type described in d). \vspace*{0.2cm} \\
{\bf f)} Now sort all of the $k$-values in increasing order, including $k_{\rm min}$ and $k_{\rm max}$. This \\
\> yields $N+1$ $k$-values which correspond to the $N+1$ amplitudes generated in b) \vspace*{0.2cm} \\
\end{tabbing}
Even though the method at first may seem confusing because it
relabels $k$-values, it in fact ensures that the $N+1$ tuple of
$k$-values is always in increasing order, but with varying
spacing. It is not important at all how the individual $k$-values
are labelled {\it while} they are generated, as long as they are
sorted in increasing order at the end.

Altogether there are now $2N+2$ numbers related to the binning and
the amplitudes. $2N$ of these are random numbers, only the $k_{\rm min}$
and $k_{\rm max}$ are fixed numbers.

The $2N$ numbers can be generated for instance by a
Monte Carlo based algorithm or they can be run on a grid.
The algorithm is straightforward to incorporate into existing
likelihood calculators such as Markov Chain Monte Carlo, it just
introduces $2N$ new ``cosmological parameters'' into the analysis.

By this method it is for instance possible in principle to sample
a step-like feature perfectly with $N=4$ (but not with N=2).

Apart from sliding bins the method is completely identical to that described
in the previous section. However, it has the very big advantage of being
able to detect narrow features of almost any shape and location
in the primordial power spectrum.

\subsection{Application to present data}

The results of likelihood analyses with $N=2,4,$ and 8 are shown
in Table \ref{table:slide}. From this it is clear that when
applied to the present cosmological data the algorithm does not
perform significantly better than the crude binning algorithm of
the previous section. This is not surprising since there are no
detectable features of any significance in the present data. For a
given $N$ the obtained $\chi^2$ is in each case slightly better
than what is obtained with the fixed bins method of the previous
section. However, the Goodness-of-Fit is actually worse because
there are $2N$ free parameters in the sliding bins method and only
$N+1$ in the fixed bins method.

\begin{figure}
\begin{center}
\includegraphics[width=110mm]{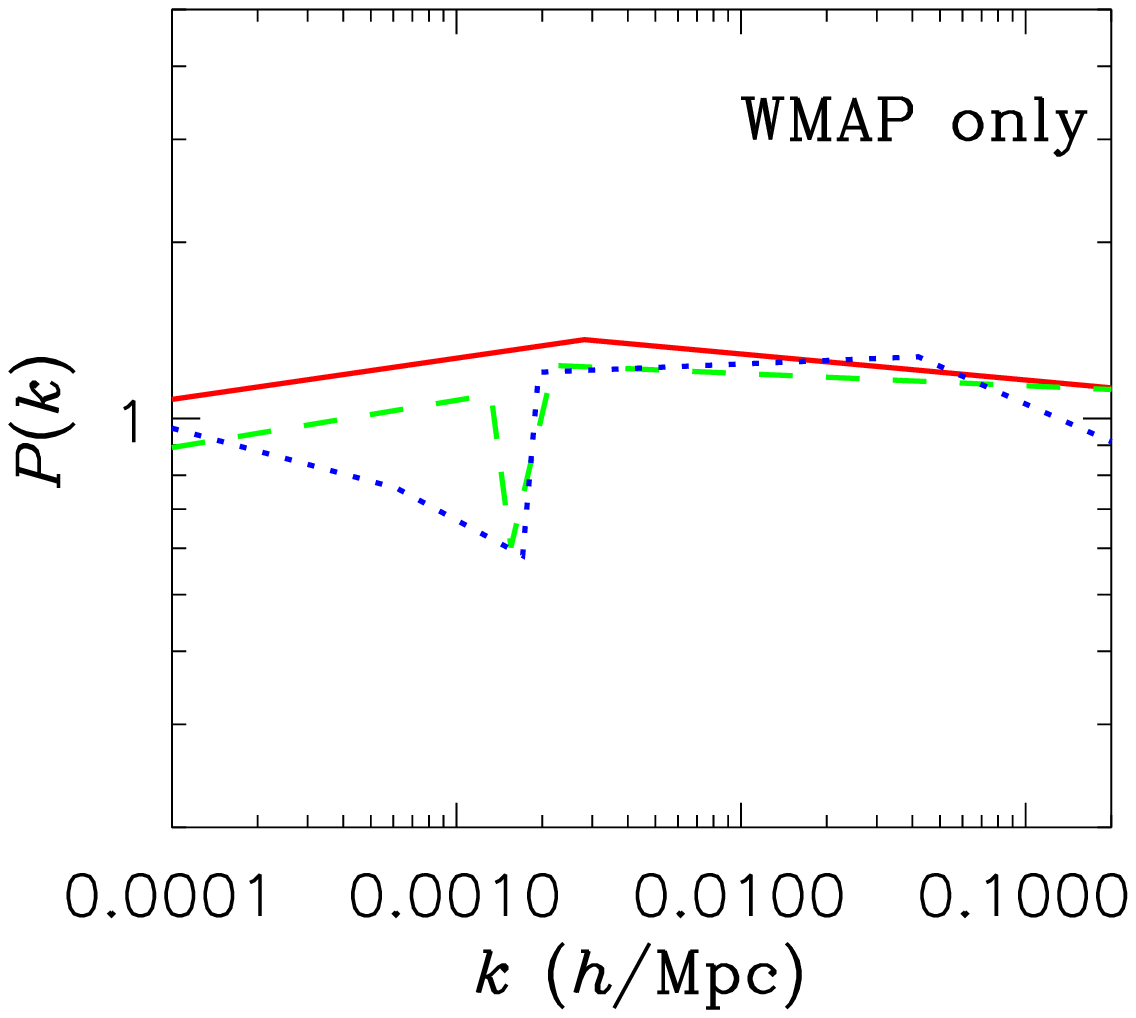}
\caption{Reconstructed best fit power spectra to the WMAP data
from the sliding bin algorithm. The full line is for $N=2$, the
dashed for $N=4$, and the dotted for $N=8$.} \label{fig:spec3}
\end{center}
\end{figure}

\begin{figure}
\begin{center}
\includegraphics[width=110mm]{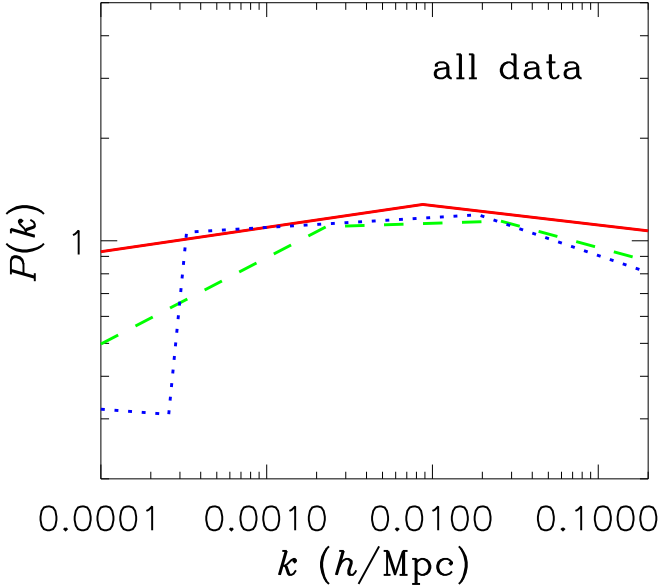}
\caption{Reconstructed best fit power spectra to all present data
from the sliding bin algorithm. The full line is for $N=2$, the
dashed for $N=4$, and the dotted for $N=8$.} \label{fig:spec4}
\end{center}
\end{figure}

Figures \ref{fig:spec3} and \ref{fig:spec4} show the reconstructed
spectra. As in the case of the algorithm with fixed binning there
is indication of an overall bending of the spectrum, corresponding
to a negative curvature. Furthermore there is indication of a dip
in power at large scales. For the WMAP-only data the dip is at $k
= 0.001-0.002 \, h$/Mpc, corresponding roughly to $l \sim 10$,
whereas for all data combined the dip occurs at a scale
corresponding to roughly $l = 2-4$.

However, as with the fixed binning algorithm there is no
statistical evidence for features in the spectrum. The GoF is
constant or decreases when $N$ increases, which indicates that
there are no significant features hidden.

One other important feature should be noted, namely that there is
apparently a rather large change in the derived power spectrum on
large scales when including other data sets than WMAP. At first
this may appear confusing because the other data sets are relevant
only on smaller scales. However, the simulated annealing algorithm
used is very powerful for finding the global extremum of a
function, but it is not an MCMC type algorithm, i.e.\ it does not
provide an estimate of the distribution of the function. Therefore
no error bars are plotted in these figures. One should instead
think of the analysis as a 1-parameter likelihood analysis, with
$N$ being the free parameter. The actual spectra plotted in the
figures are the best fit models only. Close to the horizon scale
the error bars on the extracted spectrum are large and therefore
the features appear more dramatic than they really are. Instead
one should look at the Goodness-of-fit as a function of $N$, in
which case there is no really significant evidence for features.

Even though the sliding bin method does not perform better than
the fixed bin method for the present data, it performs extremely
well when there are localized features in the spectrum, as will be
demonstrated in the next section.

\begin{table}
\begin{center}
\begin{tabular}{|l|ll|ll|}
\hline
& \multicolumn{2}{l|}{\em WMAP data only} & \multicolumn{2}{l|}{\em all data} \cr
\hline $N$ & $\chi^2$ & GoF & $\chi^2$ & GoF \cr \hline 2 &
1430.73 & 0.0421 & 1463.42 & 0.0779 \cr 4 & 1428.14 & 0.0397 &
1460.08 & 0.0759 \cr 8 & 1428.08 & 0.0282 & 1457.59 & 0.0620 \cr
\hline
\end{tabular}
\end{center}
\caption{The smallest $\chi^2$ recovered from the sliding bin
algorithm, based on the given data set. The Goodness-of-Fit (GoF)
is defined in Eq.~(\ref{eq:gof}).} \label{table:slide}
\end{table}

\section{Testing the algorithm on mock CMB data}

We have simulated a data set from the future Planck mission using
the following very simple prescription:
We assume it to be cosmic variance (as opposed to foreground) limited
up to some maximum $l$-value, $l_{\rm max}$, which we take to be
2000.
For the sake of simplicity we shall work only with the temperature
power spectrum, $C_{T,l}$, in the present analysis.
In fact the Planck detectors will be able to measure polarization
as well as temperature anisotropies, but our simplification of using
only $C_{T,l}$ will not have a significant qualitative impact on our
conclusions. For a cosmic variance limited full sky experiment the
uncertainty in the measurement of a given $C_l$ is simply
\begin{equation}
\frac{\sigma (C_l)}{C_l} = \sqrt{\frac{2}{2l+1}}.
\end{equation}
It should be noted that taking the data to be cosmic variance limited
corresponds to the best possible case. In reality it is likely
that foreground effects will be significant, especially at high
$l$. Therefore, our estimate of the
precision with which the oscillation parameters can be measured
is probably on the optimistic side.

As the underlying cosmological model for all cases in this and the
next section we choose a standard $\Lambda$CDM model with
$\Omega_0 = 1$, $\Omega_m = 0.3$, $\Omega_b = 0.04$, $H_0 = 70
\,\, {\rm km} \, {\rm s}^{-1} \, {\rm Mpc}^{-1}$, and $\tau = 0$.


\subsection{A step in the spectrum}

As a test case we use a primordial power spectrum which has a
step at $k = 0.05 h$/Mpc,
\begin{equation}
P(k) = \cases{1 & for $k < 0.05 h$/Mpc \cr 1.3 & for $k > 0.05
h$/Mpc}
\label{eq:step}
\end{equation}

\begin{figure}
\begin{center}
\includegraphics[width=110mm]{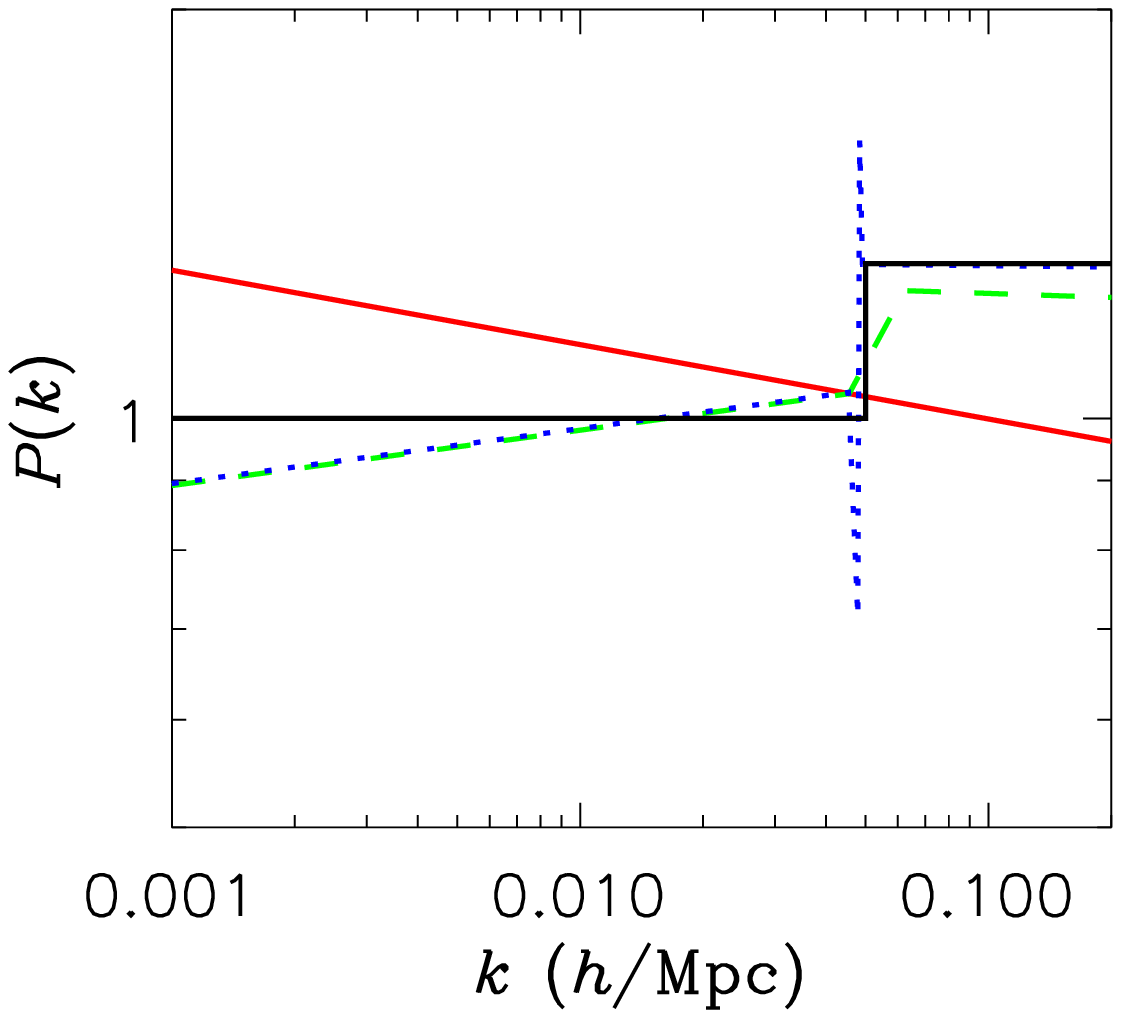}
\caption{Reconstructed best fit power spectra to the underlying
spectrum from Eq.~(\ref{eq:step}). The full line is for $N=2$, the
dashed for $N=4$, and the dotted for $N=8$. The thick (black) line
is the power spectrum of the input model.} \label{fig:step}
\end{center}
\end{figure}

In Table \ref{table:step} we show the best fit $\chi^2$ obtained
for $N=2,4,$ and 8. In this case the GoF increases dramatically
with increasing $N$, indicating that there is a feature in the
spectrum. From Figure \ref{fig:step} it can also be seen that the
step is faithfully recovered by the algorithm. For $N=4$ it is
located at the correct place in $k$-space, but is not quite sharp
enough. For $N=8$ the step feature is recovered almost perfectly,
except for a transient effect right at the edge. Such an effect is
not surprising because the algorithm has to reproduce a
discontinuity.

\begin{table}
\begin{center}
\begin{tabular}{|l|ll|}
\hline $N$ & $\chi^2$ & GoF \cr \hline 2 & 13038.78 & $7.17 \times
10^{-1590}$ \cr 4 & 2251.23 & $2.80 \times 10^{-5}$ \cr 8 &
2147.63 & 0.00442 \cr \hline
\end{tabular}
\end{center}
\caption{The smallest $\chi^2$ recovered from the model with a
step-like feature. The Goodness-of-Fit (GoF) is defined in
Eq.~(\ref{eq:gof}).} \label{table:step}
\end{table}

\subsection{A Gaussian feature}

The next case is a Gaussian suppression of the spectrum around
$k = 0.05 h$/Mpc,
\begin{equation}
P(k) = 1-\alpha
\exp\left(-\frac{(\log(k)-\log(k_b))^2}{\beta^2}\right),
\label{eq:gauss}
\end{equation}
where we have taken $k_b = 0.05 h$/Mpc, $\alpha = 0.7$, $\beta^2 =
0.1$.

In Table \ref{table:gauss} the best fit $\chi^2$ and GoF are
shown, again for $N=2,4,$ and 8. Figure 8 shows the recovered
spectra.

\begin{figure}
\begin{center}
\includegraphics[width=110mm]{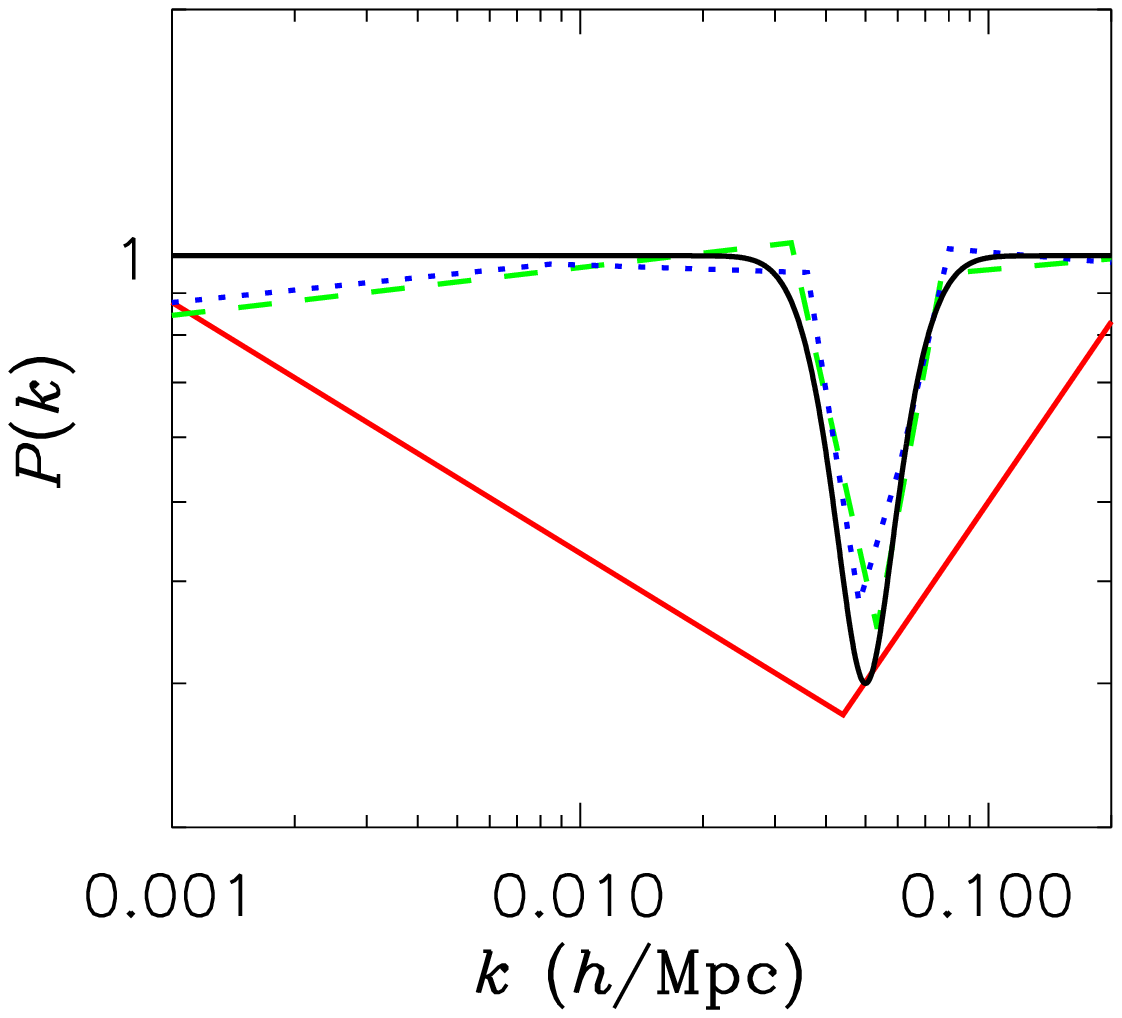}
\caption{Reconstructed best fit power spectra to the underlying
spectrum from Eq.~(\ref{eq:gauss}). The full line is for $N=2$,
the dashed for $N=4$, and the dotted for $N=8$. The thick (black)
line is the power spectrum of the input model.} \label{fig:gauss}
\end{center}
\end{figure}

\begin{table}
\begin{center}
\begin{tabular}{|l|ll|}
\hline $N$ & $\chi^2$ & GoF \cr \hline 2 & 25109.22 & $6.13 \times
10^{-3928}$ \cr 4 & 2289.51 & $2.27 \times 10^{-6}$ \cr 8 &
2147.63 & $2.82 \times 10^{-5}$ \cr \hline
\end{tabular}
\end{center}
\caption{The smallest $\chi^2$ recovered from the model with a
Gaussian feature. The Goodness-of-Fit (GoF) is defined in
Eq.~(\ref{eq:gof}).} \label{table:gauss}
\end{table}

Not surprisingly the best fit even for $N=8$ is not tremendously
good because it is inherently difficult to mimic a continuous feature
in the spectrum by a sequence of lines.

However, the recovered $\chi^2$ and the reconstructed spectrum is
much better than what a crude binning algorithm can do on the same
data \cite{Hannestad:2000pm}. Furthermore the shape of the
underlying spectrum is recovered nicely and unbiased. Finally, in
figure 9 we show CMB spectra of the underlying theoretical spectra
discussed in sections 5.1 and 5.2.

\begin{figure}
\begin{center}
\includegraphics[width=110mm]{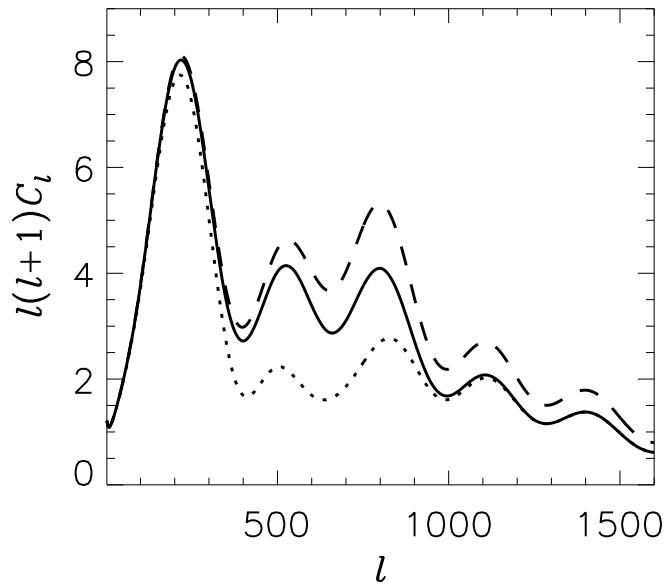}
\caption{The CMB spectra of the two simulated data sets. The
dashed line is for the model with a step, the dotted for the model
with a Gaussian feature. The full line is for a model with the
same cosmological parameters and a scale invariant spectrum.}
\label{fig:cl_feature}
\end{center}
\end{figure}


\section{Glitches in the CMB spectrum}

\begin{figure}
\begin{center}
\hspace*{1.8cm}\includegraphics[width=120mm]{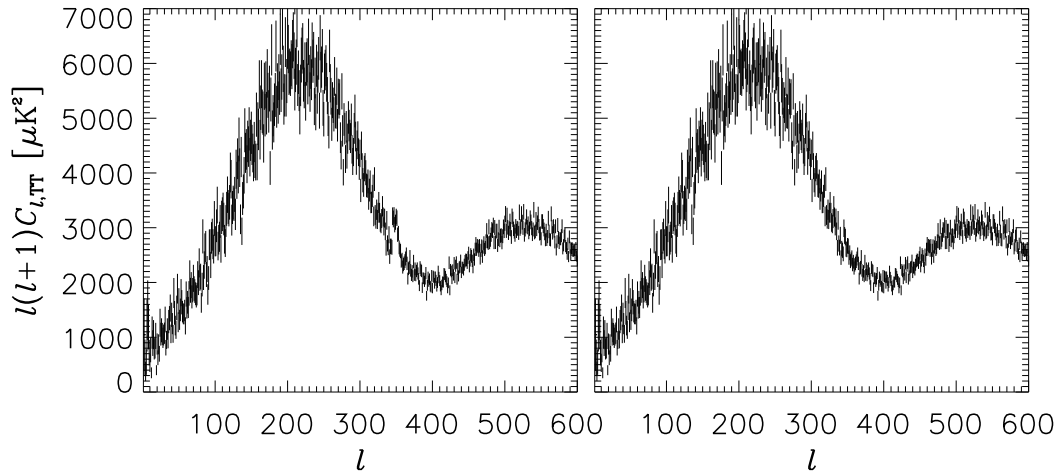} \caption{The
two sets of mock CMB data. The right figure shows the unperturbed
observational data, whereas the left curve shows the data with a
Gaussian peak added at $l=350$.} \label{fig:fake}
\end{center}
\end{figure}

As a last application we study what happens to spectrum estimation
in the event where there are significant, but completely
unphysical bumps on the $C_l$ spectrum. To do this we have
generated a mock CMB data in the same manner as in the previous
section for a completely scale invariant model ($n=1$), but on top
of that we have added a Gaussian feature, so that the final $C_l$
in given in terms of the original $C_l$ as
\begin{equation}
C_{l,{\rm new}} = C_{l,{\rm old}}(1 +0.2 \exp(-(l-350)^2/5^5))
\label{eq:fake}
\end{equation}
In figure \ref{fig:fake} we show the two sets of mock CMB data.
The left spectrum has the added Gaussian feature, while the right
one is the original unperturbed mock spectrum.

The question is then to what extent the estimation, both  of the spectrum
and of the cosmological parameters, is perturbed by this feature.

\begin{figure}
\begin{center}
\includegraphics[width=110mm]{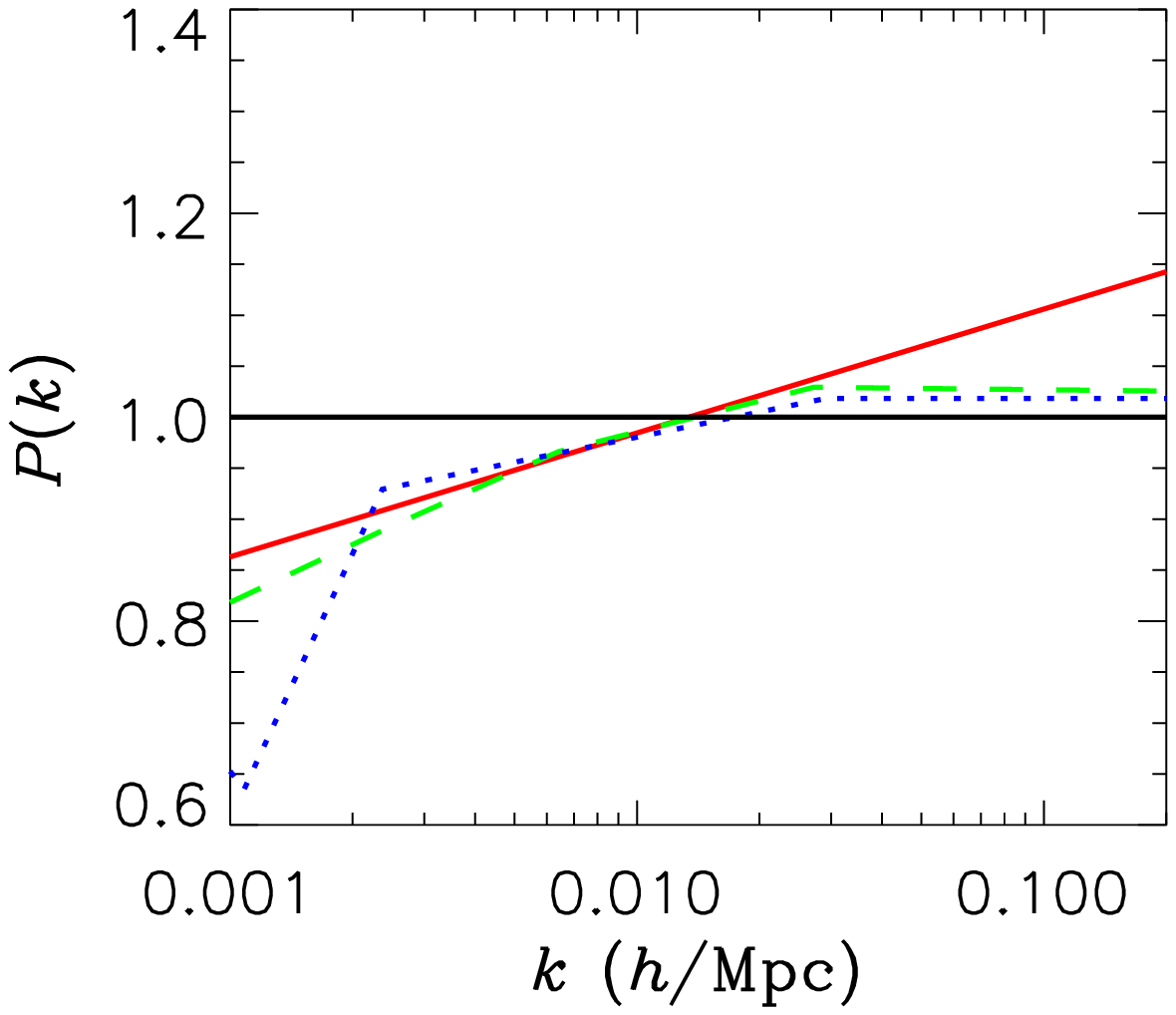}
\caption{Reconstructed best fit power spectra to the underlying
scale invariant spectrum, modified according to
Eq.~(\ref{eq:fake}). The full line is for $N=2$, the dashed for
$N=4$, and the dotted for $N=8$. The thick (black) line is the
power spectrum of the input model.} \label{fig:fake2}
\end{center}
\end{figure}

In Table \ref{table:fake} it can be seen that the true, underlying
parameters are not retrieved in the case with $N=2$. However,
already for $N=4$ the cosmological parameters are estimated
correctly. On the other hand, from figure \ref{fig:fake2} it can
be seen that the extracted power spectrum is no longer the fully
scale invariant $n=1$ Harrison-Zeldovich spectrum of the input
model. Rather, the best fit model now shows a distinct bending.
This shows that errors in the $C_l$ construction from the map in
the form of bumps or glitches can lead to a biasing of the power
spectrum estimation. Although the feature in the $C_l$ spectrum
which is used here is significantly larger than the ones seen in
the WMAP data (as can be seen from the fact that the $N=2$ model
gives an exceedingly bad fit) it is quite possible to bias power
spectrum estimation and make an underlying scale invariant
spectrum look like it is bending.

The fact that the other cosmological parameters can be extracted
with great precision is not too surprising. The reason is that
parameters such as $\Omega_m$, $H_0$, and $\Omega_b$ all produce
wide non-localized features in the spectrum. Therefore it is quite
impossible to mimic a narrow Gaussian feature by a linear
combination of changes in the underlying cosmological parameters.

\begin{table}
\begin{center}
\begin{tabular}{|l|ll|ccc|}
\hline $N$ & $\chi^2$ & GoF & $\Omega_m h^2$ & $\Omega_b h^2$ & h
\cr \hline 2 & 3425.72 & $6.97 \times 10^{-80}$ & 0.179 & 0.0203 &
0.600 \cr 4 & 2133.43 & 0.0114 & 0.146 & 0.0199 & 0.706 \cr 8 &
2131.30 & 0.00885 & 0.146 & 0.0198 & 0.707 \cr  \hline
\end{tabular}
\end{center}
\caption{The smallest $\chi^2$ recovered based on the modified
spectrum from Eq.~(\ref{eq:fake}), as well as the best fit values
of $\Omega_m h^2$, $\Omega_b h^2$, and $h$. The Goodness-of-Fit
(GoF) is defined in Eq.~(\ref{eq:gof}).} \label{table:fake}
\end{table}

\subsection{A glitch at low $l$}

Finally we have also tested the effect of having a glitch at lower
$l$ in order to see whether that produces similar effects. We have
added a glitch of the same magnitude and width as above, but at
$l=40$.

\begin{figure}
\begin{center}
\includegraphics[width=110mm]{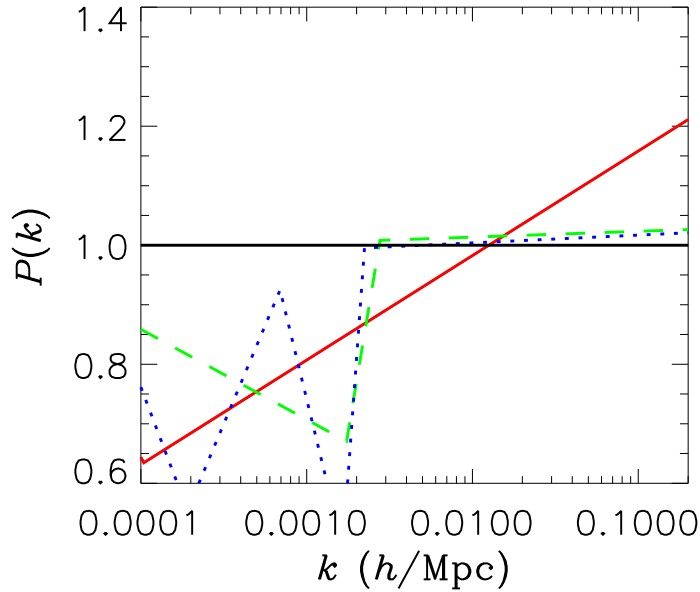}
\caption{Reconstructed best fit power spectra to the underlying
scale invariant spectrum, modified according to
Eq.~(\ref{eq:fake}), but with the feature located at $l=40$. The
full line is for $N=2$, the dashed for $N=4$, and the dotted for
$N=8$. The thick (black) line is the power spectrum of the input
model.} \label{fig:fake3}
\end{center}
\end{figure}

From figure 12 it can be seen that in this case the reconstructed
spectrum also shows features. The main effect is at $k \sim 0.001
h$/Mpc, corresponding to roughly $l \sim 40$, but here again there
are features at larger scales as well. In general narrow features
in the $C_l$ spectrum will not show up as narrow features in the
reconstructed $P(k)$.


\section{Discussion}

We have developed an efficient and model-independent method for
reconstructing the primordial power spectrum from cosmological
observations of both CMB and large scale structure. This method
has the major advantage that it can locate even narrow features in
the spectrum very accurately. Furthermore the algorithm is very
easy to implement in current likelihood calculators which are
Monte Carlo based, although it can also be used in grid based
algorithms.

We then tested the algorithm on current cosmological data and
found, in accordance with numerous other studies, no significant
evidence for any features in the primordial spectrum of
fluctuations, $P(k)$. There is a slight preference for an overall
negative curvature of the spectrum, corresponding to a negative
running of the spectral index. However, the statistical
significance is very low. Furthermore there is an indication of a
break in the power spectrum at the very largest observable scales,
corresponding to the dip in power at low $l$ seen by WMAP. Again,
however, the statistical significance of this feature is very low.

In order to test the efficiency and robustness of the algorithm we
then went on to use it on mock data from a future high-precision
experiment such as the Planck Surveyor. Two cases were
investigated, one with a step-like feature imposed, and one with a
Gaussian suppression of the power spectrum. In both cases the
algorithm was able to recover the true, underlying power spectrum
very efficiently and with no biasing, something which would for
instance be extremely difficult for algorithms with fixed bins.

Finally we tested the algorithm on a mock CMB spectrum, where a
noise feature of Gaussian shape was imposed {\it after} the data
was generated. The reason is that such glitches are seen in the
present WMAP data, and it is of interest to know whether the true
power spectrum, as well as other cosmological parameters can be
safely recovered even if there are such spurious features present.

Running the algorithm we found that the cosmological parameters of
the underlying model could be recovered safely, but that the
extracted primordial spectrum no longer resembled to
scale-invariant spectrum used to generate the data. Rather it
showed a distinct bending. This means that care should be taken
when extracting information on the primordial power spectrum,
including such parameters as the running of the spectral index,
from power spectra which show spurious, localized features in
$l$-space.

\section*{Acknowledgments}

We acknowledge use of the publicly available CMBFAST package
written by Uros Seljak and Matthias Zaldarriaga \cite{CMBFAST}
and the use of computing resources at DCSC (Danish Center for
Scientific Computing).

\pagebreak

\newcommand\AJ[3]{~Astron. J.{\bf ~#1}, #2~(#3)}
\newcommand\APJ[3]{~Astrophys. J.{\bf ~#1}, #2~ (#3)}
\newcommand\apjl[3]{~Astrophys. J. Lett. {\bf ~#1}, L#2~(#3)}
\newcommand\ass[3]{~Astrophys. Space Sci.{\bf ~#1}, #2~(#3)}
\newcommand\cqg[3]{~Class. Quant. Grav.{\bf ~#1}, #2~(#3)}
\newcommand\mnras[3]{~Mon. Not. R. Astron. Soc.{\bf ~#1}, #2~(#3)}
\newcommand\mpla[3]{~Mod. Phys. Lett. A{\bf ~#1}, #2~(#3)}
\newcommand\npb[3]{~Nucl. Phys. B{\bf ~#1}, #2~(#3)}
\newcommand\plb[3]{~Phys. Lett. B{\bf ~#1}, #2~(#3)}
\newcommand\pr[3]{~Phys. Rev.{\bf ~#1}, #2~(#3)}
\newcommand\prog[3]{~Prog. Theor. Phys.{\bf ~#1}, #2~(#3)}

\end{document}